\documentclass[11pt]{article}
\usepackage[a4paper, top=2.0cm, bottom=2.0cm, left=2.0cm, right=2.0cm]{geometry}
\usepackage{multicol}
\usepackage{setspace}
\usepackage[scaled]{helvet}
 
\usepackage[T1]{fontenc}
\usepackage{graphicx}
\usepackage{amssymb}
\usepackage{epstopdf}
\usepackage{wrapfig}
\usepackage{color}
\usepackage{chemformula}
\usepackage{caption}
\usepackage{upgreek}
\usepackage[dvipsnames]{xcolor}
\usepackage[font={color=black},figurename=Fig.,labelfont={it}]{caption}
\usepackage{siunitx}
\usepackage{authblk}
\usepackage{hyperref}
\usepackage{xcolor}
\usepackage{enumitem} 
\usepackage{paralist}
\setitemize{noitemsep,topsep=0pt,parsep=0pt,partopsep=0pt,leftmargin=1.5\parindent}
\hypersetup{
colorlinks=true,
linkcolor=blue,
urlcolor=blue,
citecolor=blue,
}
\usepackage{natbib}
\usepackage{aasmacros}
\usepackage{doi}
\usepackage[labelformat=empty]{caption}
\usepackage{soul}

\newcommand{\squishlist}{
 \begin{list}{$\bullet$}
  { \setlength{\itemsep}{1pt}
     \setlength{\parsep}{0pt}
     \setlength{\topsep}{1pt}
     \setlength{\partopsep}{0pt}
     \setlength{\leftmargin}{1.5em}
     \setlength{\labelwidth}{1.5em}
     \setlength{\labelsep}{0.5em} } }
\newcommand{\squishend}{
  \end{list}  }

\begin{document} 

\title{The Key to Unlocking Exoplanet Biosignatures: \\ a UK-led IR Spectrograph for the Habitable Worlds Observatory Coronagraph (theme: Astro)}
\author{co-lead authors (co-authors and signatories at end): \\ Beth Biller, University of Edinburgh, \\ Dan Dicken, UK Astronomy Technology Centre \\}

\date{}

\maketitle

\section{Executive Summary}

The detection of life on rocky exoplanets in the habitable zones of nearby stars would be a  paradigm-shifting advance, and it is one of the greatest scientific challenges of our time. There is no single spectral feature that is an unambiguous sign of life on a given exoplanet. Instead, the current state-of-the-art approach involves detecting multiple molecular atmospheric features that should not exist together in equilibrium, e.g. simultaneous detection of O$_2$ and CH$_4$. Spectra across a wide wavelength (0.3-1.7 $\mu$m) range are necessary to cover multiple spectral features per molecule of interest and to  contextualise the suite of molecular features detected. While the US will lead the optical arm of the Habitable Worlds Observatory (HWO) coronagraph, a UK-led contribution of a near-infrared Integral Field Spectrograph (IFS) for the infrared arm will ensure UK leadership in the flagship scientific goal of HWO - to search for signatures of life on potentially habitable exoplanets.

\section{Scientific Motivation \& Objectives}

Whether rocky exoplanets in the habitable zones of nearby stars harbour life is one of the most important open scientific questions of our lifetime.
Both the ESA Voyage 2050 roadmap and the NASA 2020 Decadal Review place detection and characterisation of temperate exoplanets as a top priority, towards the ultimate goal of remote detection of extraterrestrial life. The discovery of signatures of life on exoplanets would be a paradigm-shifting advance, but is not possible with the current generation of telescopes and instruments, except for rocky planets orbiting low-mass stars. However,  rocky planets around low-mass stars are blasted with X-rays and gamma rays from their host stars, producing a hostile radiation environment compared to the Sun. It is not clear whether these planets are habitable or if they can retain atmospheres \citep{Zieba2023, Ducrot2025, Piaulet-Ghorayeb2025}. Instead, the best place to search for “life as we know it” are planets similar in mass to our own that orbit stars of similar mass to our Sun (FGK stars). 

The NASA-led Habitable Worlds Observatory (HWO) will search these worlds for signs of life while opening a new era in astronomy as an ultraviolet, optical, and near-infrared space telescope, purpose-built to succeed Hubble and extend humanity’s understanding of the cosmos. With a 6 to 8 meter primary mirror positioned at the Sun–Earth L2 point alongside JWST, HWO will deliver unprecedented performance and capability at UV through NIR wavelengths. At the core of the HWO mission, the Coronagraph Instrument (CI) will unlock the technologies needed to discover and characterise Earth-like worlds in the habitable zones of solar-type stars, setting the stage for humanity’s first steps toward finding life beyond our planet. Only high-contrast imaging from space-based platforms can achieve the 10$^{-10}$ planet-star contrast necessary to detect and characterise such planets \citep{Currie2023}. 

\subsection{The Importance of NIR Coverage}
\begin{figure}
    \includegraphics[width=1.0\textwidth]{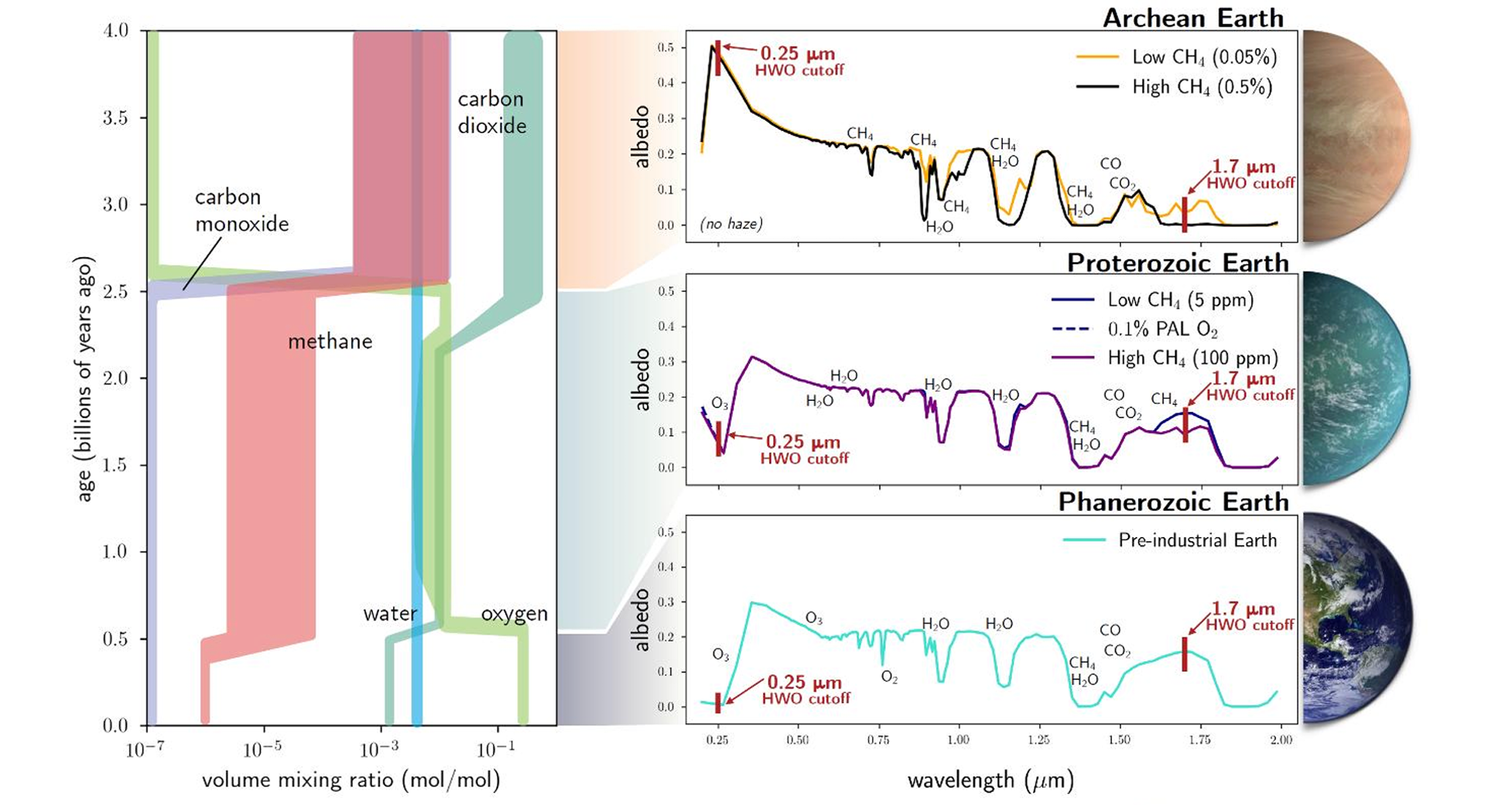} \caption{Figure 1: Earth's spectrum over time (right panel), reproduced from \cite{Krissansen-Totton2025}.  The biosignature gases present change dramatically from epoch to epoch (left panel), producing significantly different spectra.  The simultaneous detection of O$_2$ at $<$0.8 $\mu$m 
    with H$_2$O, CH$_4$, and CO$_2$ at $>$0.8 $\mu$m demonstrates the presence of life in the modern Earth spectrum shown in the bottom right panel, but for habitable planets without much evidence of oxygenic biosignatures (e.g. Proterozoic and Archean Earth), coverage beyond 0.8 $\mu$m will be key to capture H$_2$O, and CH$_4$ features.} \label{earth_over_time}
\end{figure}

The primary mission of HWO is to identify and directly image  potentially habitable worlds, to determine the frequency of habitable zone rocky planets around stars like our own Sun and detect atmospheric biosignatures. There is no single spectral feature that is an unambiguous sign of life on a given exoplanet; instead, the current state-of-the-art approach involves detecting multiple molecular atmospheric features that should not exist together in equilibrium, e.g. simultaneous detection of O$_2$, H$_2$O, CH$_4$, and CO$_2$ \citep{schwieterman2018, Schwieterman2024}.  
Thus, spectra across a wide wavelength (0.3-1.7 $\mu$m) range are necessary to cover multiple spectral features per molecule of interest and to fully contextualise the suite of molecular features detected, in order to rule out false positives for any particular molecular feature \citep{damiano2022, Krissansen-Totton2025}.  We show a simulated spectrum of the modern Earth in the bottom right panel of Figure~\ref{earth_over_time}; the UV-optical arm of the spectrograph will cover O$_3$ and O$_2$ features, but coverage into the near-infrared is necessary to also probe water and carbon dioxide features, which are required for a positive ID of the presence of life.  Habitable exoplanets also may not necessarily resemble the modern Earth, given that Earth's spectral features have changed dramatically over time. The modern Earth atmosphere has only existed for 10\% of Earth's inhabited history over the last 4 billion years. Before the advent of oxygenic biosignatures on Earth, most biosignature gas features appeared at wavelengths $>$0.8 $\mu$m (Figure~\ref{earth_over_time}), underscoring the need for IR spectroscopy to detect biosignatures for the full range of potentially habitable exoplanets.   


The HWO Coronagraph Instrument will be specifically designed to obtain reflected-light imaging and low-resolution spectroscopy of habitable zone rocky planets over a wavelength range of 0.3 to 1.7 $\mu$m by obscuring the host star, yielding both the first detections of habitable zone rocky planets around FGK star hosts and the potential to search for indications of biosignature gases in their atmospheres.  With both a UV-optical (0.3-0.8 $\mu$m) and a near-IR (0.8-1.7 $\mu$m) arm, the HWO CI will require custom-built integral field spectrographs for each arm, including dedicated focal plane detector systems specific to each wavelength range. The integral field spectrograph (IFS) captures spatially resolved spectra across a two-dimensional field, enabling simultaneous imaging across the entire waveband alongside detailed spectral analysis of every point in the scene. This advanced spectrograph is essential for multiple reasons: it provides the capability to characterise and suppress speckle patterns from residual starlight subtraction, thereby improving achievable contrast; it enables simultaneous spectral capture of multiple planets orbiting the same host star; and it supports coronagraphic science on extended structures such as circumstellar disks.

\subsection{Beyond Habitable Rocky Worlds -- Broader Science Impacts}
The Coronagraph Instrument (CI) will deliver transformative science well beyond its core exoplanet mission. By suppressing light from bright central objects, CI opens new frontiers in astrophysics. Space-based coronagraphs on HST and JWST have already revolutionized circumstellar disk studies; HWO’s coronagraph will push further with an unprecedented inner working angle, probing disk regions at rocky-planet scales.
This capability will also enable breakthrough AGN research, revealing how jets interact with galactic gas and dust, regulate star formation, and shape galaxy evolution \cite{Zhang2025}.
The Solar System provides compelling science cases for the near-IR IFS coronagraphic channel. Small bodies—asteroids, Centaurs, and Trans-Neptunian Objects are relics of early planetesimal formation. Coronagraphic imaging will resolve their immediate environments at high contrast, enabling searches for satellites, rings, and jets from cryovolcanic activity. These observations will reveal their dynamical evolution, internal structure, and composition, addressing fundamental questions about their origin and the processes that shaped them. CI will also deliver transformational advances in detecting and characterizing a broad range of exoplanets - not only rocky, habitable worlds but also widely separated planets that do not transit their stars \cite{Sagynbayeva2025, Renyu2025}. This includes solar system analogues of gas and ice giants, as well as super-Earths absent from our own planetary system, dramatically expanding our understanding of planetary atmospheres and diversity. 
These capabilities build on foundations laid by coronagraphs on the Nancy Grace Roman Space Telescope and METIS on the ELT. METIS will host the first infrared integral field spectrograph (notably UK-led and UK-built) behind coronagraph optics on a flagship facility, positioning the UK at the forefront of this revolution.


\section{Strategic Context}

HWO is NASA's next flagship mission, poised to deliver transformational impacts across multiple fields of astrophysics. The UK community has a strong heritage in such missions, most notably through leadership of the Mid-Infrared Instrument (MIRI) for JWST \citep{Wright2023}. Importantly, MIRI was not part of JWST's original instrument suite; it emerged as a community-driven proposal that extended JWST's wavelength coverage into the mid-infrared. Its resounding success underscores the value of international leadership and collaboration. Building on this legacy, the UK is now uniquely positioned to contribute to the \emph{primary} instrument suite of HWO. The UK currently has the unprecedented opportunity to participate directly in HWO's core mission and science goal — encoded in the observatory's name — the search for and detection of biosignatures on rocky, habitable-zone exoplanets.
\textbf{An in-kind UK contribution to the Coronagraph Instrument would secure direct involvement and leadership in achieving HWO's flagship objective—and potentially enable one of the most profound scientific discoveries of the 21st century.}

We strongly advocate for the UK community to lead the development and construction of a near-infrared spectrograph for the HWO Coronagraph Instrument. While NASA will likely lead the UV/optical arm of the coronagraph, discussions with Dr.~Feng Zhao (HWO Deputy Chief Technologist, JPL) and the HWO project office at NASA Goddard confirm that international contributions are actively sought for much of the near-infrared arm. Although groups in the US, Japan, and Europe have expressed interest in building the IR coronagraph optics, the UK is currently the only community proposing to deliver an infrared spectrograph backend—an essential component to unlock the full scientific potential of the instrument. This effort plays directly to UK strengths, leveraging our heritage in integral field spectrographs for MIRI on JWST and current instruments such as METIS for the ELT.

The proposed IFS subsystem would also house the near-infrared detector focal plane—an opportunity to showcase UK leadership in space detector technology. Leonardo UK’s APD detectors set the benchmark for near-noiseless photon counting, a capability essential for detecting faint signals from solar system analogue exoplanets. This technology aligns closely with ESA’s GAIA-NIR development track, reinforcing the UK’s strategic position in next-generation missions.

Moreover, this UK ambition aligns fully with the broader goals of the European high-contrast imaging and spectroscopy community, which is highly organised and active \cite{Laginja2025}. Germany, France and Italy bring deep expertise from missions such as MIRI on JWST \cite{Wright2023}, the Roman Coronagraph \cite{Poberezhskiy2025}, and SPHERE on the VLT \cite{Chomez2025}, as well as leadership in next-generation ELT instruments like METIS \cite{Delacroix2024, Feldt2024}, ANDES \cite{Palle2025}, HARMONI \cite{Vaughan2024} and PCS 
\cite{Kasper2021}. Over the past two years, this wider community has convened two major workshops with more than 50 participants \cite{Laginja2025}, focused on advancing research and development for space-based high-contrast imaging missions—particularly HWO and the Large Interferometer for Exoplanets (LIFE). A key outcome from the May 2025 workshop was the urgent need to accelerate the development of specialised cryogenic vacuum testbeds in Europe to raise the technology readiness level of critical near-IR coronagraphic technologies. Such a facility would complement NASA’s JPL testbed — used to mature the visible-channel coronagraph for HWO based on Roman heritage — and provide essential infrastructure for testing a UK-built IFS. Preparations are also underway to bid for Horizon Europe Synergy funding to advance high-contrast imaging technologies for HWO at a European scale.

\section{Proposed Approach and Technical Solution}

The HWO Coronagraph Instrument will include at least two channels, each operating as an independent system. These channels are designed to observe simultaneously, with incoming light from the exoplanet system spectrally and spatially separated by a high-performance dichroic filter. At the core of the coronagraphs are advanced nulling optics that suppress starlight to reveal exoplanets that can be up to ten billion times fainter than their host star. Multiple technology tracks are under investigation to achieve this nulling for HWO \cite{McElwain2025}. Once the starlight is suppressed, the next critical step is to feed the signal into the backend integral field spectrographs (IFS).

This is where the UK contribution offers a unique advantage: the input to the IFS is the nulled image, independent of telescope design or nulling technology. The near-infrared IFS and focal plane system can be developed as a modular subsystem—a “plug-in” component for the coronagraph. This approach allows us to advance IFS development now, without depending on the outcome of other technology tracks and telescope design changes and tradeoffs, enabling us to quickly mature the technologies required for the spectrograph and detector system. We plan to finalize the IFS design and identify technology gaps that must be addressed to achieve a Technology Readiness Level (TRL) of at least 5 by the end of this decade (a NASA requirement), coinciding with HWO’s Mission Concept Review (MCR). TRL 5 will demonstrate that the technology performs under conditions similar to those expected in the mission environment, though not yet in the full operational environment e.g., thermal, vacuum, vibration for space hardware. Key technologies on this roadmap include: \\

\textbullet~~\textbf{Detectors:} photon-counting and noiseless readout systems \\

\textbullet~~\textbf{High-performance optics:} high-throughput \& low-scatter dichroics, filters, and

~~~microlens/microslicer arrays essential for IFS functionality \\

\textbullet~~\textbf{Precision mechanisms:} ultra-stable, repeatable systems for mode and band switching  \\


All these technologies are currently at TRL~3 or higher, making it realistic to advance their maturity, validate their suitability, and deliver meaningful contributions to HWO ahead of MCR. This strategy will secure UK leadership in the project and establish us as a trusted partner in one of the most ambitious astrophysics missions of the century.

Engaging early in the coronagraph architecture—well before final concept design—is key to establishing UK leadership in the HWO mission with NASA. This aligns with NASA’s strategic vision to mature architecture, technology, and science holistically and early, building in system-level robustness from the start. Unlike other potential HWO instruments still seeking definition, the coronagraph is already confirmed as central to the observatory’s mission to detect habitable worlds—enshrined in both the programme’s science goals and name. This enables us to begin targeted design and technology development now, giving a substantial head start over competing concepts and ensuring meaningful contributions well before MCR at the end of the decade. Early-stage architecture and design requirements are already being explored with NASA, including material choices, spectral resolution, and field of view.
 
\section{UK Leadership and Capability}

The scientific field of exoplanets is one of the youngest fields in Astrophysics, but also one of the highest impact scientific research fields, reflected in the award of the Nobel prize in 2019 for the first discovery of exoplanets.  The transformational impact of this field has been recognised in both the ESA Voyage 2050 roadmap and the NASA 2020 Decadal Review, which put detection and characterisation of temperate exoplanets as the highest priority scientific goal of the coming decades.  The UK has a strong and growing exoplanet community, with research groups (at least one permanent member of academic staff) at: Imperial College London, Keele University, King’s College London, Queen Mary University of London, Queen’s University Belfast, The Open University, UCL, Birmingham, Bristol, Cambridge, Cardiff, Central Lancashire, Edinburgh, Exeter, Leeds, Leicester, Hertfordshire, Manchester, Oxford, St Andrews, Warwick.  
The yearly UK Exoplanet Meeting has regularly drawn 150-200 participants since 2014.  The UK community has particular strengths in characterisation of exoplanet atmospheres via both transit and direct imaging techniques (e.g. \cite{Ahrer2023, Miles2023}), as well as in high-precision radial velocity studies (e.g. \cite{Lienhard2025}).

\begin{figure}
    \includegraphics[width=1.0\textwidth]{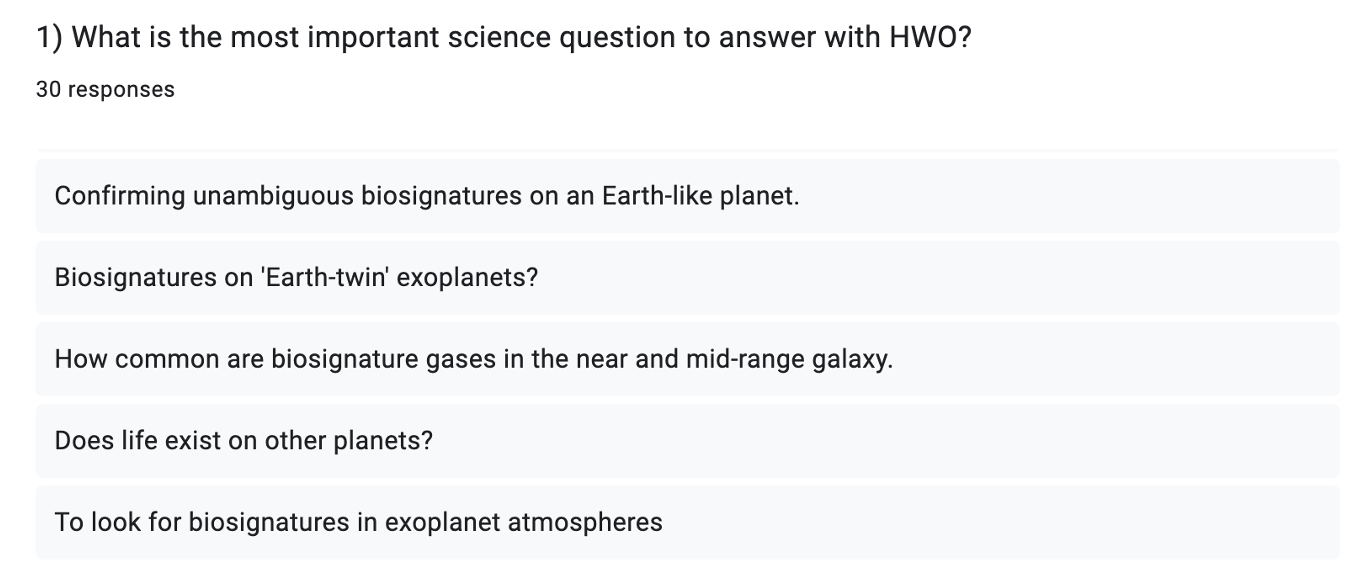} \caption{Figure 2: An online survey of the UK exoplanet community shows overwhelming consensus: the highest scientific priority for HWO is detecting biosignatures on rocky exoplanets in the habitable zones of Sun-like stars.  More details on this survey available here: \url{http://www.exocommunity.uk/networks/HWO/HWO.html}. } \label{onlinesurvey}
\end{figure}

The UK exoplanet community firmly recognises HWO, and in particular, the flagship case of detection and characterisation of rocky, habitable zone exoplanets, as the critical next step in exoplanet science.  In Autumn 2024, the UK exoplanet community was polled to determine the interests and priorities of this community with respect to HWO.  The goal of the survey was to find areas of consensus across the community to identify scientific / technical areas where the UK could provide a substantial national contribution and thus serve as a basis for organisation of the community.  The community overwhelmingly identified biosignature detection on rocky exoplanets in the habitable zones of Sunlike stars as their top scientific question with HWO (see Fig.~\ref{onlinesurvey} for sample responses).  At the 2025 UK Exoplanet Meeting in April 2025 in Leeds, the community was given an additional chance to respond via an interactive poster.  Respondents indicated that they strongly back a UK in-kind contribution to HWO.  Given that the community identifies biosignature detection as the top science question with HWO and the strong support for a UK in-kind contribution to HWO, contributing a spectrograph for the CI would offer the highest impact for the community and the best chance of leadership in the crucial scientific case of biosignature detection.

Successful detection of biosignatures in exoplanets will require more than just hardware contributions, but also expertise in observational techniques for high-contrast imaging as well as significant effort in modelling of exoplanet atmospheres in order to enable robust identification of biosignatures.  The UK leads in this field, with notable expertise in atmospheric characterisation of both transiting and directly imaged exoplanets with JWST, e.g. \cite{Ahrer2023, Miles2023}, as well as in atmospheric modelling of habitable exoplanets, e.g. \cite{Rugheimer2018, Braam2022}.  Support of an IFS for the CI would further boost these efforts in the UK.

The UK ATC is positioned at the forefront of this initiative, leveraging its heritage, experience, and strong partnerships established through leadership on JWST. With Europe’s largest astronomy optics group and flagship projects such as LISA, METIS, and HARMONI in the pipeline, ATC is exceptionally well placed to lead this study. However, a project of this scale demands a broad and highly skilled team. We envision close collaboration with UK partners, including RAL Space, which played a central role in MIRI development, and leading academic institutions such as Edinburgh, Durham, Cardiff, UCL, Exeter, Oxford, Open University and Cambridge, all of which bring deep expertise in optics, detector systems, and space missions.

A key work package in the initial study will identify potential UK hardware contributions, with Leonardo UK emerging as the clear industry front-runner for detector technology. Beyond the focal plane system, other critical elements—such as optical filters, precision surfaces, and mechanisms may be supplied by UK industry. Furthermore, the UK’s strength in testing infrastructure and Assembly, Integration, and Test (AIT) for space missions is well established, with trusted teams already supporting numerous ESA and NASA projects. This capability will be essential for advancing technology readiness and ensuring the UK plays a leading role in delivering this flagship mission.

\section{Partnership Opportunities}



Contributing to the HWO Coronagraph Instrument offers a timely and strategic opportunity to strengthen UK collaboration with NASA’s leading centers — JPL and Goddard Space Flight Center — both central to HWO. JPL is advancing coronagraph development, while Goddard hosts the project office and technology testbeds. The UK has a proven track record with NASA through leadership on JWST’s MIRI instrument and support during its critical commissioning phase. However, more than a decade has passed since MIRI’s delivery, making HWO the ideal mission to renew and expand these trusted partnerships and engage a new generation of UK engineers and scientists.

This contribution will also reinforce UK leadership within Europe and position us to shape a broader European role in HWO. A strong UK-NASA partnership will enhance our influence on ESA’s strategy and foster collaboration across Europe. Recent workshops with European high-contrast imaging groups and NASA have already begun defining this path. By advancing development of the near-infrared integral field spectrograph (IFS) for the coronagraph, the UK can lead a critical technical area, complementing expertise in France and Germany, where institutions like MPIA have collaborated with JPL on Roman’s coronagraph. Beyond Europe, UK ATC has initiated discussions with JAXA, which has expressed interest in NIR coronagraph optics, with both JAXA and European partners seeing no barriers to future collaboration. These partnerships align with the UK’s industrial strategy and strengthen our global position in flagship missions.
 
\section{Conclusion}

Developing an infrared integral field spectrograph for the Coronagraph Instrument aligns perfectly with UK expertise, built on a distinguished track record in infrared instrumentation—including leadership of the MIRI IFS on JWST and the METIS IFS and HARMONI instruments for the ELT. By applying this heritage to deliver a world-leading spectrograph for HWO, the UK community currently has the unprecedented
opportunity to play a leading role in HWO’s core science goal — the search for and detection of biosignatures on rocky, habitable-zone exoplanets.

\newpage

\noindent \textbf{Co-authors:} \\
Olivier Absil (Liege) \\
Raziye Artan (UK ATC) \\
Jo Barstow (Open University) \\
Jayne Birkby (University of Oxford) \\
Christophe Dumas (Director, UK ATC) \\
Sasha Hinkley (University of Exeter) \\
Tad Komacek (University of Oxford) \\
Katherine Morris (UK ATC) \\
Lorenzo Pino (INAF - Osservatorio Astrofisico di Arcetri) \\
Sarah Rugheimer (University of Edinburgh) \\
Colin Snodgrass (University of Edinburgh) \\
Stephen Todd (UK ATC) \\
Vinooja Thurairethinam (UK ATC) \\
Amaury Triaud (University of Birmingham) \\

\noindent \textbf{Co-signers:} \\
Suzanne Aigrain (University of Oxford) \\
Amirnezam Amiri ( University of Arkansas) \\
Mariangela Bonavita (University of Edinburgh) \\
Mark Booth (UK ATC) \\
Richard Booth (University of Leeds) \\
Marrick Braam (University of Bern) \\
Matteo Brogi (University of Turin) \\
David J. A. Brown (University of Warwick) \\
Andrew Collier Cameron (University of St Andrews) \\
\'Oscar Carri\'on-Gonz\'alez (Max Planck Institute for Astronomy) \\
Gael Chauvin (Max Planck Institute for Astronomy) \\
Xueqing Chen (University of Edinburgh) \\
Katy L. Chubb (University of Bristol) \\
Alastair Claringbold (University of Warwick) \\
Matthew Cole (University of Edinburgh) \\
Angaraj Duara (Centre for Electronic Imaging, Open University) \\
Trent Dupuy (University of Edinburgh) \\
Cl\'emence Fontanive (University of Edinburgh) \\
Gergely Friss (University of Edinburgh) \\
Edward Gillen (Queen Mary University of London) \\
Carole Haswell (The Open University) \\
Tanja Holc (University of Edinburgh) \\
Aiza Kenzhebekova (University of Edinburgh) \\
Adam Koval (University of Edinburgh) \\
James Kirk (Imperial College London) \\
David A. Lewis (University of Edinburgh) \\
Nikku Madhusudhan (University of Cambridge) \\
Mei Ting Mak (University of Oxford) \\
Subhanjoy Mohanty (Imperial College London) \\
Vincent Okoth (University of Edinburgh) \\
Larissa Palethorpe (University of Edinburgh) \\
Paul Palmer (University of Edinburgh) \\
Vatsal Panwar (University of Birmingham) \\
Mia Belle Parkinson (University of Edinburgh) \\
Chris Pearson (RAL Space) \\
Anjali Piette (University of Birmingham) \\
Ken Rice (University of Edinburgh) \\
Subhajit Sarakr (Cardiff University) \\
Adam Stevenson (University of Birmingham) \\
Ben Sutlieff (University of Edinburgh) \\
Jonathan Tennyson (University College London) \\
Eleni Tsiakaliari (Open University UK) \\
Daniel Valentine (University of Bristol) \\
Joost P. Wardenier (University of Bern) \\

\bibliographystyle{abbrv}
\bibliography{bib}

@ARTICLE{Zieba2023,
       author = {{Zieba}, Sebastian and {Kreidberg}, Laura and {Ducrot}, Elsa and {Gillon}, Micha{\"e}l and {Morley}, Caroline and {Schaefer}, Laura and {Tamburo}, Patrick and {Koll}, Daniel D.~B. and {Lyu}, Xintong and {Acu{\~n}a}, Lorena and {Agol}, Eric and {Iyer}, Aishwarya R. and {Hu}, Renyu and {Lincowski}, Andrew P. and {Meadows}, Victoria S. and {Selsis}, Franck and {Bolmont}, Emeline and {Mandell}, Avi M. and {Suissa}, Gabrielle},
        title = "{No thick carbon dioxide atmosphere on the rocky exoplanet TRAPPIST-1 c}",
      journal = {\nat},
         year = 2023,
        month = aug,
       volume = {620},
       number = {7975},
        pages = {746-749},
          doi = {10.1038/s41586-023-06232-z},
archivePrefix = {arXiv},
       eprint = {2306.10150},
 primaryClass = {astro-ph.EP},
       adsurl = {https://ui.adsabs.harvard.edu/abs/2023Natur.620..746Z}
}

@ARTICLE{Ducrot2025,
       author = {{Ducrot}, Elsa and {Lagage}, Pierre-Olivier and {Min}, Michiel and {Gillon}, Micha{\"e}l and {Bell}, Taylor J. and {Tremblin}, Pascal and {Greene}, Thomas and {Dyrek}, Achr{\`e}ne and {Bouwman}, Jeroen and {Waters}, Rens and {G{\"u}del}, Manuel and {Henning}, Thomas and {Vandenbussche}, Bart and {Absil}, Olivier and {Barrado}, David and {Boccaletti}, Anthony and {Coulais}, Alain and {Decin}, Leen and {Edwards}, Billy and {Gastaud}, Ren{\'e} and {Glasse}, Alistair and {Kendrew}, Sarah and {Olofsson}, Goran and {Patapis}, Polychronis and {Pye}, John and {Rouan}, Daniel and {Whiteford}, Niall and {Argyriou}, Ioannis and {Cossou}, Christophe and {Glauser}, Adrian M. and {Krause}, Oliver and {Lahuis}, Fred and {Royer}, Pierre and {Scheithauer}, Silvia and {Colina}, Luis and {van Dishoeck}, Ewine F. and {Ostlin}, G{\"o}ran and {Ray}, Tom P. and {Wright}, Gillian},
        title = "{Combined analysis of the 12.8 and 15 {\ensuremath{\mu}}m JWST/MIRI eclipse observations of TRAPPIST-1 b}",
      journal = {Nature Astronomy},
         year = 2025,
        month = mar,
       volume = {9},
        pages = {358-369},
          doi = {10.1038/s41550-024-02428-z},
archivePrefix = {arXiv},
       eprint = {2412.11627},
 primaryClass = {astro-ph.EP},
       adsurl = {https://ui.adsabs.harvard.edu/abs/2025NatAs...9..358D}
}

@ARTICLE{Piaulet-Ghorayeb2025,
       author = {{Piaulet-Ghorayeb}, Caroline and {Benneke}, Bj{\"o}rn and {Turbet}, Martin and {Moore}, Keavin and {Roy}, Pierre-Alexis and {Lim}, Olivia and {Doyon}, Ren{\'e} and {Fauchez}, Thomas J. and {Albert}, Lo{\"\i}c and {Radica}, Michael and {Coulombe}, Louis-Philippe and {Lafreni{\`e}re}, David and {Cowan}, Nicolas B. and {Belzile}, Danika and {Musfirat}, Kamrul and {Kaur}, Mehramat and {L'Heureux}, Alexandrine and {Johnstone}, Doug and {MacDonald}, Ryan J. and {Allart}, Romain and {Dang}, Lisa and {Kaltenegger}, Lisa and {Pelletier}, Stefan and {Rowe}, Jason F. and {Taylor}, Jake and {Turner}, Jake D.},
        title = "{Strict Limits on Potential Secondary Atmospheres on the Temperate Rocky Exo-Earth TRAPPIST-1 d}",
      journal = {\apj},
         year = 2025,
        month = aug,
       volume = {989},
       number = {2},
          eid = {181},
        pages = {181},
          doi = {10.3847/1538-4357/adf207},
archivePrefix = {arXiv},
       eprint = {2508.08416},
 primaryClass = {astro-ph.EP},
       adsurl = {https://ui.adsabs.harvard.edu/abs/2025ApJ...989..181P}
}

@ARTICLE{Rugheimer2018,
       author = {{Rugheimer}, S. and {Kaltenegger}, L.},
        title = "{Spectra of Earth-like Planets through Geological Evolution around FGKM Stars}",
      journal = {\apj},
         year = 2018,
        month = feb,
       volume = {854},
       number = {1},
          eid = {19},
        pages = {19},
          doi = {10.3847/1538-4357/aaa47a},
archivePrefix = {arXiv},
       eprint = {1712.10027},
 primaryClass = {astro-ph.EP},
       adsurl = {https://ui.adsabs.harvard.edu/abs/2018ApJ...854...19R}
}

@ARTICLE{Braam2022,
       author = {{Braam}, Marrick and {Palmer}, Paul I. and {Decin}, Leen and {Ridgway}, Robert J. and {Zamyatina}, Maria and {Mayne}, Nathan J. and {Sergeev}, Denis E. and {Abraham}, N. Luke},
        title = "{Lightning-induced chemistry on tidally-locked Earth-like exoplanets}",
      journal = {\mnras},
         year = 2022,
        month = dec,
       volume = {517},
       number = {2},
        pages = {2383-2402},
          doi = {10.1093/mnras/stac2722},
archivePrefix = {arXiv},
       eprint = {2209.12502},
 primaryClass = {astro-ph.EP},
       adsurl = {https://ui.adsabs.harvard.edu/abs/2022MNRAS.517.2383B}
}

@ARTICLE{Lienhard2025,
       author = {{Lienhard}, F. and {Mortier}, A. and {Cameron}, A. Collier and {Cretignier}, M. and {Borsato}, L. and {John}, A. Anna and {Egger}, J.~A. and {Stalport}, M. and {Wilson}, T.~G. and {Deline}, A. and {Fortier}, A. and {Latham}, D.~W. and {Malavolta}, L. and {Maxted}, P.~F.~L. and {Sousa}, S.~G. and {Grimm}, S.~L. and {Buchhave}, L. and {Alibert}, Y. and {Lakeland}, B.~S. and {Dumusque}, X. and {Cabrera}, J. and {Naponiello}, L. and {Correia}, A.~C.~M. and {Rescigno}, F. and {Fossati}, L. and {Sozzetti}, A. and {Alonso}, R. and {B{\'a}rczy}, T. and {Barrado}, D. and {Barros}, S.~C.~C. and {Baumjohann}, W. and {Benz}, W. and {Billot}, N. and {Brandeker}, A. and {Broeg}, C. and {Collins}, K. and {Csizmadia}, Sz and {Cubillos}, P.~E. and {Davies}, M.~B. and {Deleuil}, M. and {Demangeon}, O.~D.~S. and {Demory}, B.-O. and {Derekas}, A. and {Edwards}, B. and {Ehrenreich}, D. and {Erikson}, A. and {Fridlund}, M. and {Gandolfi}, D. and {Gazeas}, K. and {Gillon}, M. and {G{\"u}del}, M. and {G{\"u}nther}, M.~N. and {Haywood}, R. and {Heitzmann}, A. and {Helling}, Ch and {Isaak}, K.~G. and {Jenkins}, J.~M. and {Kiss}, L.~L. and {Korth}, J. and {Lam}, K.~W.~F. and {Laskar}, J. and {Etangs}, A. Lecavelier des and {Leleu}, A. and {Lendl}, M. and {Magrin}, D. and {Mart{\'\i}nez Fiorenzano}, A.~F. and {Mer{\'\i}n}, B. and {Mordasini}, C. and {Nascimbeni}, V. and {Olofsson}, G. and {Osborn}, H.~P. and {Ottensamer}, R. and {Pagano}, I. and {Palethorpe}, L. and {Pall{\'e}}, E. and {Peter}, G. and {Piazza}, D. and {Piotto}, G. and {Pollacco}, D. and {Queloz}, D. and {Ragazzoni}, R. and {Rando}, N. and {Rauer}, H. and {Ribas}, I. and {Rice}, K. and {Santos}, N.~C. and {Scandariato}, G. and {S{\'e}gransan}, D. and {Simon}, A.~E. and {Smith}, A.~M.~S. and {Sulis}, S. and {Szab{\'o}}, Gy M. and {Udry}, S. and {Ulmer-Moll}, S. and {Van Grootel}, V. and {Venturini}, J. and {Villaver}, E. and {Walton}, N.~A. and {Zingales}, T.},
        title = "{HARPS-N, TESS, and CHEOPS<SUP></SUP> discover a transiting sub-Neptune and two outer companions around the bright solar analogue HD 85426}",
      journal = {\mnras},
         year = 2025,
        month = nov,
          doi = {10.1093/mnras/staf1934},
archivePrefix = {arXiv},
       eprint = {2511.08473},
 primaryClass = {astro-ph.EP},
       adsurl = {https://ui.adsabs.harvard.edu/abs/2025MNRAS.tmp.1875L}
}

@ARTICLE{Kasper2021,
       author = {{Kasper}, M. and {Cerpa Urra}, N. and {Pathak}, P. and {Bonse}, M. and {Nousiainen}, J. and {Engler}, B. and {Heritier}, C.~T. and {Kammerer}, J. and {Leveratto}, S. and {Rajani}, C. and {Bristow}, P. and {Le Louarn}, M. and {Madec}, P.-Y. and {Str{\"o}bele}, S. and {Verinaud}, C. and {Glauser}, A. and {Quanz}, S.~P. and {Helin}, T. and {Keller}, C. and {Snik}, F. and {Boccaletti}, A. and {Chauvin}, G. and {Mouillet}, D. and {Kulcs{\'a}r}, C. and {Raynaud}, H.-F.},
        title = "{PCS {\textemdash} A Roadmap for Exoearth Imaging with the ELT}",
      journal = {The Messenger},
         year = 2021,
        month = mar,
       volume = {182},
        pages = {38-43},
          doi = {10.18727/0722-6691/5221},
archivePrefix = {arXiv},
       eprint = {2103.11196},
 primaryClass = {astro-ph.IM},
       adsurl = {https://ui.adsabs.harvard.edu/abs/2021Msngr.182...38K}
}

@ARTICLE{Zhang2025,
       author = {{Zhang}, Lulu and {Kaur}, Gagandeep and {Gao}, Tianmu and {Labiano}, {\'A}lvaro and {Hicks}, Erin K.~S. and {U}, Vivian and {Packham}, Chris and {Mehdipour}, Missagh and {Fischer}, Travis and {Storchi Bergmann}, Thaisa and {Roy}, Namrata and {M{\'a}rquez}, Isabel and {Boersma}, Christiaan},
        title = "{Deciphering The Launching of Multi-phase AGN-driven Outflows and Their (Spatially Resolved) Multi-scale Impact}",
      journal = {arXiv e-prints},
         year = 2025,
        month = aug,
          eid = {arXiv:2508.01437},
        pages = {arXiv:2508.01437},
          doi = {10.48550/arXiv.2508.01437},
archivePrefix = {arXiv},
       eprint = {2508.01437},
 primaryClass = {astro-ph.GA},
       adsurl = {https://ui.adsabs.harvard.edu/abs/2025arXiv250801437Z}
}

@ARTICLE{Renyu2025,
       author = {{Hu}, Renyu and {Min}, Michiel and {Millar-Blanchaer}, Max and {Lustig-Yaeger}, Jacob and {Robinson}, Tyler and {Burt}, Jennifer and {Coustenis}, Athena and {Damiano}, Mario and {Dong}, Chuanfei and {Dressing}, Courtney and {Fossati}, Luca and {Kane}, Stephen and {Kelkar}, Soumil and {Lichtenberg}, Tim and {Ruffio}, Jean-Baptiste and {Sur}, Dibyendu and {Tokadjian}, Armen and {Turbet}, Martin},
        title = "{Identifying rocky planets and water worlds among sub-Neptune-sized exoplanets with the Habitable Worlds Observatory}",
      journal = {arXiv e-prints},
         year = 2025,
        month = sep,
          eid = {arXiv:2509.16798},
        pages = {arXiv:2509.16798},
          doi = {10.48550/arXiv.2509.16798},
archivePrefix = {arXiv},
       eprint = {2509.16798},
 primaryClass = {astro-ph.IM},
       adsurl = {https://ui.adsabs.harvard.edu/abs/2025arXiv250916798H}
}

@ARTICLE{Sagynbayeva2025,
       author = {{Sagynbayeva}, Sabina and {Abbas}, Asif and {Kane}, Stephen R. and {Nielsen}, Eric L. and {Thompson}, William and {Blunt}, Sarah and {Rice}, Malena and {Christiansen}, Jessie L. and {Harada}, Caleb K. and {Newton}, Elisabeth R. and {Hasegawa}, Yasuhiro and {Armitage}, Philip J. and {Daylan}, Tansu},
        title = "{Requirements for Joint Orbital Characterization of Cold Giants and Habitable Worlds with Habitable Worlds Observatory}",
      journal = {\aj},
         year = 2025,
        month = oct,
       volume = {170},
       number = {4},
          eid = {208},
        pages = {208},
          doi = {10.3847/1538-3881/adf84d},
archivePrefix = {arXiv},
       eprint = {2507.21443},
 primaryClass = {astro-ph.EP},
       adsurl = {https://ui.adsabs.harvard.edu/abs/2025AJ....170..208S}
}

@ARTICLE{Poberezhskiy2025,
       author = {{Poberezhskiy}, Ilya and {Cady}, Eric and {Heydorff}, Kathryn and {Kern}, Brian and {Luchik}, Thomas and {Zhao}, Feng and {Bailey}, Vanessa and {Bush}, Nathan and {Colavita}, Mark and {Creager}, Brandon and {Fathpour}, Nanaz and {Gaidon}, Clement and {Grue}, Amanda and {Kempenaar}, Joshua and {Krist}, John and {Kuan}, Gary and {Lam}, Jonathan and {Mandi{\'c}}, Milan and {Marx}, David and {Nemati}, Bijan and {Eldorado Riggs}, A.~J. and {Seo}, Byoung-Joon and {Shi}, Fang and {Smith}, Matthew W. and {Zhou}, Hanying},
        title = "{Overview of Roman Coronagraph Instrument requirements, test campaign, and results}",
      journal = {Journal of Astronomical Telescopes, Instruments, and Systems},
         year = 2025,
        month = jul,
       volume = {11},
          eid = {031511},
        pages = {031511},
          doi = {10.1117/1.JATIS.11.3.031511},
       adsurl = {https://ui.adsabs.harvard.edu/abs/2025JATIS..11c1511P}
}

@ARTICLE{Chomez2025,
       author = {{Chomez}, A. and {Delorme}, P. and {Lagrange}, A.-M. and {Gratton}, R. and {Flasseur}, O. and {Chauvin}, G. and {Langlois}, M. and {Mazoyer}, J. and {Zurlo}, A. and {Desidera}, S. and {Mesa}, D. and {Bonnefoy}, M. and {Feldt}, M. and {Hagelberg}, J. and {Meyer}, M. and {Vigan}, A. and {Ginski}, C. and {Kenworthy}, M. and {Albert}, D. and {Bergeon}, S. and {Beuzit}, J.-L. and {Biller}, B. and {Bhowmik}, T. and {Boccaletti}, A. and {Bonavita}, M. and {Brandner}, W. and {Cantalloube}, F. and {Cheetham}, A. and {D'Orazi}, V. and {Dominik}, C. and {Fontanive}, C. and {Galicher}, R. and {Henning}, Th. and {Janson}, M. and {Kral}, Q. and {Lagadec}, E. and {Lazzoni}, C. and {Le Coroller}, H. and {Ligi}, R. and {Maire}, A.-L. and {Marleau}, G.-D. and {Menard}, F. and {Messina}, S. and {Meunier}, N. and {Mordasini}, C. and {Moutou}, C. and {M{\"u}ller}, A. and {Perrot}, C. and {Samland}, M. and {Schmid}, H.~M. and {Schmidt}, T. and {Squicciarini}, V. and {Sissa}, E. and {Turatto}, M. and {Udry}, S. and {Abe}, L. and {Antichi}, J. and {Asensio-Torres}, R. and {Baruffolo}, A. and {Baudoz}, P. and {Baudrand}, J. and {Bazzon}, A. and {Blanchard}, P. and {Bohn}, A.~J. and {Brown Sevilla}, S. and {Carbillet}, M. and {Carle}, M. and {Cascone}, E. and {Charton}, J. and {Claudi}, R. and {Costille}, A. and {De Caprio}, V. and {Delboulb{\'e}}, A. and {Dohlen}, K. and {Engler}, N. and {Fantinel}, D. and {Feautrier}, P. and {Fusco}, T. and {Gigan}, P. and {Girard}, J.~H. and {Giro}, E. and {Gisler}, D. and {Gluck}, L. and {Gry}, C. and {Hubin}, N. and {Hugot}, E. and {Jaquet}, M. and {Kasper}, M. and {Le Mignant}, D. and {Llored}, M. and {Madec}, F. and {Magnard}, Y. and {Martinez}, P. and {Maurel}, D. and {M{\"o}ller-Nilsson}, O. and {Mouillet}, D. and {Moulin}, T. and {Orign{\'e}}, A. and {Pavlov}, A. and {Perret}, D. and {Petit}, C. and {Pragt}, J. and {Puget}, P. and {Rabou}, P. and {Ramos}, J. and {Rickman}, E.~L. and {Rigal}, F. and {Rochat}, S. and {Roelfsema}, R. and {Rousset}, G. and {Roux}, A. and {Salasnich}, B. and {Sauvage}, J.-F. and {Sevin}, A. and {Soenke}, C. and {Stadler}, E. and {Suarez}, M. and {Wahhaj}, Z. and {Weber}, L. and {Wildi}, F.},
        title = "{The SPHERE infrared survey for exoplanets (SHINE): IV. Complete observations, data reduction and analysis, detection performances, and final results}",
      journal = {\aap},
         year = 2025,
        month = may,
       volume = {697},
          eid = {A99},
        pages = {A99},
          doi = {10.1051/0004-6361/202451751},
archivePrefix = {arXiv},
       eprint = {2501.12002},
 primaryClass = {astro-ph.EP},
       adsurl = {https://ui.adsabs.harvard.edu/abs/2025A&A...697A..99C}
}

@ARTICLE{Vaughan2024,
       author = {{Vaughan}, Sophia R. and {Birkby}, Jayne L. and {Thatte}, Niranjan and {Carlotti}, Alexis and {Houll{\'e}}, Mathis and {Pereira-Santaella}, Miguel and {Clarke}, Fraser and {Vigan}, Arthur and {Lin}, Zifan and {Kaltenegger}, Lisa},
        title = "{Behind the mask: can HARMONI@ELT detect biosignatures in the reflected light of Proxima b?}",
      journal = {\mnras},
         year = 2024,
        month = feb,
       volume = {528},
       number = {2},
        pages = {3509-3522},
          doi = {10.1093/mnras/stae242},
archivePrefix = {arXiv},
       eprint = {2401.09589},
 primaryClass = {astro-ph.EP},
       adsurl = {https://ui.adsabs.harvard.edu/abs/2024MNRAS.528.3509V}
}

@ARTICLE{Palle2025,
       author = {{Palle}, Enric and {Biazzo}, Katia and {Bolmont}, Emeline and {Molli{\`e}re}, Paul and {Poppenhaeger}, Katja and {Birkby}, Jayne and {Brogi}, Matteo and {Chauvin}, Gael and {Chiavassa}, Andrea and {Hoeijmakers}, Jens and {Lellouch}, Emmanuel and {Lovis}, Christophe and {Maiolino}, Roberto and {Nortmann}, Lisa and {Parviainen}, Hannu and {Pino}, Lorenzo and {Turbet}, Martin and {Weder}, Jesse and {Albrecht}, Simon and {Antoniucci}, Simone and {Barros}, Susana C. and {Beaudoin}, Andre and {Benneke}, Bjorn and {Boisse}, Isabelle and {Bonomo}, Aldo S. and {Borsa}, Francesco and {Brandeker}, Alexis and {Brandner}, Wolfgang and {Buchhave}, Lars A. and {Cheffot}, Anne-Laure and {Deborde}, Robin and {Debras}, Florian and {Doyon}, Rene and {Di Marcantonio}, Paolo and {Giacobbe}, Paolo and {Gonz{\'a}lez Hern{\'a}ndez}, Jonay I. and {Helled}, Ravit and {Kreidberg}, Laura and {Machado}, Pedro and {Maldonado}, Jesus and {Marconi}, Alessandro and {Martins}, B.~L. Canto and {Miceli}, Adriano and {Mordasini}, Christoph and {N'Diaye}, Mamadou and {Niedzielski}, Andrzej and {Nisini}, Brunella and {Origlia}, Livia and {Peroux}, Celine and {Pietrow}, Alexander G.~M. and {Pinna}, Enrico and {Rauscher}, Emily and {Reffert}, Sabine and {Rodr{\'\i}guez-L{\'o}pez}, Cristina and {Rousselot}, Philippe and {Sanna}, Nicoletta and {Santos}, Nuno C. and {Simonnin}, Adrien and {Su{\'a}rez Mascare{\~n}o}, Alejandro and {Zanutta}, Alessio and {Zapatero-Osorio}, Maria Rosa and {Zechmeister}, Mathias},
        title = "{Ground-breaking exoplanet science with the ANDES spectrograph at the ELT}",
      journal = {Experimental Astronomy},
         year = 2025,
        month = jun,
       volume = {59},
       number = {3},
          eid = {29},
        pages = {29},
          doi = {10.1007/s10686-025-10000-4},
archivePrefix = {arXiv},
       eprint = {2311.17075},
 primaryClass = {astro-ph.IM},
       adsurl = {https://ui.adsabs.harvard.edu/abs/2025ExA....59...29P}
}

@INPROCEEDINGS{Delacroix2024,
       author = {{Delacroix}, Christian and {K{\"o}nig}, Lorenzo and {Absil}, Olivier and {Orban De Xivry}, Gilles and {Forsberg}, Pontus and {Karlsson}, Mikael and {Ronayette}, Samuel and {Pantin}, Eric and {Barri{\`e}re}, Jean-Christophe},
        title = "{The ELT/METIS annular groove phase masks}",
    booktitle = {Advances in Optical and Mechanical Technologies for Telescopes and Instrumentation VI},
         year = 2024,
       editor = {{Navarro}, Ram{\'o}n and {Jedamzik}, Ralf},
       series = {Society of Photo-Optical Instrumentation Engineers (SPIE) Conference Series},
       volume = {13100},
        month = aug,
          eid = {131002R},
        pages = {131002R},
          doi = {10.1117/12.3020539},
       adsurl = {https://ui.adsabs.harvard.edu/abs/2024SPIE13100E..2RD}
}

@ARTICLE{Feldt2024,
       author = {{Feldt}, Markus and {Bertram}, Thomas and {Correia}, Carlos and {Absil}, Olivier and {C{\'a}rdenas V{\'a}zquez}, M. Concepci{\'o}n and {Coppejans}, Hugo and {Kulas}, Martin and {Obereder}, Andreas and {Orban de Xivry}, Gilles and {Scheithauer}, Silvia and {Steuer}, Horst},
        title = "{High strehl and high contrast for the ELT instrument METIS: Final design, implementation, and predicted performance of the single-conjugate adaptive optics system}",
      journal = {Experimental Astronomy},
         year = 2024,
        month = dec,
       volume = {58},
       number = {3},
          eid = {20},
        pages = {20},
          doi = {10.1007/s10686-024-09968-2},
archivePrefix = {arXiv},
       eprint = {2411.17341},
 primaryClass = {astro-ph.IM},
       adsurl = {https://ui.adsabs.harvard.edu/abs/2024ExA....58...20F}
}

@ARTICLE{Laginja2025,
       author = {{Laginja}, Iva and {Carri{\'o}n-Gonz{\'a}lez}, {\'O}scar and {Laugier}, Romain and {Matthews}, Elisabeth and {Leboulleux}, Lucie and {Potier}, Axel and {Lau}, Alexis and {Absil}, Olivier and {Baudoz}, Pierre and {Biller}, Beth and {Boccaletti}, Anthony and {Brandner}, Wolfgang and {Carlotti}, Alexis and {Chauvin}, Ga{\"e}l and {Choquet}, {\'E}lodie and {Doelman}, David and {Dohlen}, Kjetil and {Ferrari}, Marc and {Hinkley}, Sasha and {Huby}, Elsa and {Karlsson}, Mikael and {Krause}, Oliver and {K{\"u}hn}, Jonas and {Le Duigou}, Jean-Michel and {Mazoyer}, Johan and {Mesa}, Dino and {Min}, Michiel and {Mouillet}, David and {Mugnier}, Laurent M. and {Orban de Xivry}, Gilles and {Snik}, Frans and {Vassallo}, Daniele and {Vigan}, Arthur and {de Visser}, Pieter},
        title = "{Advancing European high-contrast imaging R\&D towards the Habitable Worlds Observatory}",
      journal = {\apss},
         year = 2025,
        month = mar,
       volume = {370},
       number = {3},
          eid = {29},
        pages = {29},
          doi = {10.1007/s10509-025-04417-8},
archivePrefix = {arXiv},
       eprint = {2503.12707},
 primaryClass = {astro-ph.IM},
       adsurl = {https://ui.adsabs.harvard.edu/abs/2025Ap&SS.370...29L}
}

@ARTICLE{McElwain2025,
       author = {{McElwain}, Michael W. and {Mawet}, Dimitri and {Ruffio}, Jean-Baptiste and {Juanola Parramon}, Roser and {Lawson}, Kellen and {Le Coroller}, Herv{\'e} and {Marois}, Christian and {Millar-Blanchaer}, Max and {Nemati}, Bijan and {Redmond}, Susan and {Ren}, Bin and {Pueyo}, Laurent and {Stark}, Christopher and {Will}, Scott},
        title = "{Habitable World Discovery and Characterization: Coronagraph Concept of Operations and Data Post-Processing}",
      journal = {arXiv e-prints},
         year = 2025,
        month = oct,
          eid = {arXiv:2510.02547},
        pages = {arXiv:2510.02547},
          doi = {10.48550/arXiv.2510.02547},
archivePrefix = {arXiv},
       eprint = {2510.02547},
 primaryClass = {astro-ph.IM},
       adsurl = {https://ui.adsabs.harvard.edu/abs/2025arXiv251002547M}
}

@ARTICLE{Miles2023,
       author = {{Miles}, Brittany E. and {Biller}, Beth A. and {Patapis}, Polychronis and {Worthen}, Kadin and {Rickman}, Emily and {Hoch}, Kielan K.~W. and {Skemer}, Andrew and {Perrin}, Marshall D. and {Whiteford}, Niall and {Chen}, Christine H. and {Sargent}, B. and {Mukherjee}, Sagnick and {Morley}, Caroline V. and {Moran}, Sarah E. and {Bonnefoy}, Mickael and {Petrus}, Simon and {Carter}, Aarynn L. and {Choquet}, Elodie and {Hinkley}, Sasha and {Ward-Duong}, Kimberly and {Leisenring}, Jarron M. and {Millar-Blanchaer}, Maxwell A. and {Pueyo}, Laurent and {Ray}, Shrishmoy and {Sallum}, Steph and {Stapelfeldt}, Karl R. and {Stone}, Jordan M. and {Wang}, Jason J. and {Absil}, Olivier and {Balmer}, William O. and {Boccaletti}, Anthony and {Bonavita}, Mariangela and {Booth}, Mark and {Bowler}, Brendan P. and {Chauvin}, Gael and {Christiaens}, Valentin and {Currie}, Thayne and {Danielski}, Camilla and {Fortney}, Jonathan J. and {Girard}, Julien H. and {Grady}, Carol A. and {Greenbaum}, Alexandra Z. and {Henning}, Thomas and {Hines}, Dean C. and {Janson}, Markus and {Kalas}, Paul and {Kammerer}, Jens and {Kennedy}, Grant M. and {Kenworthy}, Matthew A. and {Kervella}, Pierre and {Lagage}, Pierre-Olivier and {Lew}, Ben W.~P. and {Liu}, Michael C. and {Macintosh}, Bruce and {Marino}, Sebastian and {Marley}, Mark S. and {Marois}, Christian and {Matthews}, Elisabeth C. and {Matthews}, Brenda C. and {Mawet}, Dimitri and {McElwain}, Michael W. and {Metchev}, Stanimir and {Meyer}, Michael R. and {Molliere}, Paul and {Pantin}, Eric and {Quirrenbach}, Andreas and {Rebollido}, Isabel and {Ren}, Bin B. and {Schneider}, Glenn and {Vasist}, Malavika and {Wyatt}, Mark C. and {Zhou}, Yifan and {Briesemeister}, Zackery W. and {Bryan}, Marta L. and {Calissendorff}, Per and {Cantalloube}, Faustine and {Cugno}, Gabriele and {De Furio}, Matthew and {Dupuy}, Trent J. and {Factor}, Samuel M. and {Faherty}, Jacqueline K. and {Fitzgerald}, Michael P. and {Franson}, Kyle and {Gonzales}, Eileen C. and {Hood}, Callie E. and {Howe}, Alex R. and {Kraus}, Adam L. and {Kuzuhara}, Masayuki and {Lagrange}, Anne-Marie and {Lawson}, Kellen and {Lazzoni}, Cecilia and {Liu}, Pengyu and {Llop-Sayson}, Jorge and {Lloyd}, James P. and {Martinez}, Raquel A. and {Mazoyer}, Johan and {Quanz}, Sascha P. and {Redai}, Jea Adams and {Samland}, Matthias and {Schlieder}, Joshua E. and {Tamura}, Motohide and {Tan}, Xianyu and {Uyama}, Taichi and {Vigan}, Arthur and {Vos}, Johanna M. and {Wagner}, Kevin and {Wolff}, Schuyler G. and {Ygouf}, Marie and {Zhang}, Xi and {Zhang}, Keming and {Zhang}, Zhoujian},
        title = "{The JWST Early-release Science Program for Direct Observations of Exoplanetary Systems II: A 1 to 20 {\ensuremath{\mu}}m Spectrum of the Planetary-mass Companion VHS 1256-1257 b}",
      journal = {\apjl},
         year = 2023,
        month = mar,
       volume = {946},
       number = {1},
          eid = {L6},
        pages = {L6},
          doi = {10.3847/2041-8213/acb04a},
archivePrefix = {arXiv},
       eprint = {2209.00620},
 primaryClass = {astro-ph.EP},
       adsurl = {https://ui.adsabs.harvard.edu/abs/2023ApJ...946L...6M}
}

@ARTICLE{Ahrer2023,
       author = {{Ahrer}, Eva-Maria and {Stevenson}, Kevin B. and {Mansfield}, Megan and {Moran}, Sarah E. and {Brande}, Jonathan and {Morello}, Giuseppe and {Murray}, Catriona A. and {Nikolov}, Nikolay K. and {Petit dit de la Roche}, Dominique J.~M. and {Schlawin}, Everett and {Wheatley}, Peter J. and {Zieba}, Sebastian and {Batalha}, Natasha E. and {Damiano}, Mario and {Goyal}, Jayesh M. and {Lendl}, Monika and {Lothringer}, Joshua D. and {Mukherjee}, Sagnick and {Ohno}, Kazumasa and {Batalha}, Natalie M. and {Battley}, Matthew P. and {Bean}, Jacob L. and {Beatty}, Thomas G. and {Benneke}, Bj{\"o}rn and {Berta-Thompson}, Zachory K. and {Carter}, Aarynn L. and {Cubillos}, Patricio E. and {Daylan}, Tansu and {Espinoza}, N{\'e}stor and {Gao}, Peter and {Gibson}, Neale P. and {Gill}, Samuel and {Harrington}, Joseph and {Hu}, Renyu and {Kreidberg}, Laura and {Lewis}, Nikole K. and {Line}, Michael R. and {L{\'o}pez-Morales}, Mercedes and {Parmentier}, Vivien and {Powell}, Diana K. and {Sing}, David K. and {Tsai}, Shang-Min and {Wakeford}, Hannah R. and {Welbanks}, Luis and {Alam}, Munazza K. and {Alderson}, Lili and {Allen}, Natalie H. and {Anderson}, David R. and {Barstow}, Joanna K. and {Bayliss}, Daniel and {Bell}, Taylor J. and {Blecic}, Jasmina and {Bryant}, Edward M. and {Burleigh}, Matthew R. and {Carone}, Ludmila and {Casewell}, S.~L. and {Changeat}, Quentin and {Chubb}, Katy L. and {Crossfield}, Ian J.~M. and {Crouzet}, Nicolas and {Decin}, Leen and {D{\'e}sert}, Jean-Michel and {Feinstein}, Adina D. and {Flagg}, Laura and {Fortney}, Jonathan J. and {Gizis}, John E. and {Heng}, Kevin and {Iro}, Nicolas and {Kempton}, Eliza M.-R. and {Kendrew}, Sarah and {Kirk}, James and {Knutson}, Heather A. and {Komacek}, Thaddeus D. and {Lagage}, Pierre-Olivier and {Leconte}, J{\'e}r{\'e}my and {Lustig-Yaeger}, Jacob and {MacDonald}, Ryan J. and {Mancini}, Luigi and {May}, E.~M. and {Mayne}, N.~J. and {Miguel}, Yamila and {Mikal-Evans}, Thomas and {Molaverdikhani}, Karan and {Palle}, Enric and {Piaulet}, Caroline and {Rackham}, Benjamin V. and {Redfield}, Seth and {Rogers}, Laura K. and {Roy}, Pierre-Alexis and {Rustamkulov}, Zafar and {Shkolnik}, Evgenya L. and {Sotzen}, Kristin S. and {Taylor}, Jake and {Tremblin}, P. and {Tucker}, Gregory S. and {Turner}, Jake D. and {de Val-Borro}, Miguel and {Venot}, Olivia and {Zhang}, Xi},
        title = "{Early Release Science of the exoplanet WASP-39b with JWST NIRCam}",
      journal = {\nat},
         year = 2023,
        month = feb,
       volume = {614},
       number = {7949},
        pages = {653-658},
          doi = {10.1038/s41586-022-05590-4},
archivePrefix = {arXiv},
       eprint = {2211.10489},
 primaryClass = {astro-ph.EP},
       adsurl = {https://ui.adsabs.harvard.edu/abs/2023Natur.614..653A}
}

@INPROCEEDINGS{Currie2023,
       author = {{Currie}, T. and {Biller}, B. and {Lagrange}, A. and {Marois}, C. and {Guyon}, O. and {Nielsen}, E.~L. and {Bonnefoy}, M. and {De Rosa}, R.~J.},
        title = "{Direct Imaging and Spectroscopy of Extrasolar Planets}",
    booktitle = {Protostars and Planets VII},
         year = 2023,
       editor = {{Inutsuka}, S. and {Aikawa}, Y. and {Muto}, T. and {Tomida}, K. and {Tamura}, M.},
       series = {Astronomical Society of the Pacific Conference Series},
       volume = {534},
        month = jul,
        pages = {799},
          doi = {10.48550/arXiv.2205.05696},
archivePrefix = {arXiv},
       eprint = {2205.05696},
 primaryClass = {astro-ph.EP},
       adsurl = {https://ui.adsabs.harvard.edu/abs/2023ASPC..534..799C}
}

\end{document}